# Multi-space excitation as an alternative to the Landauer picture for non-equilibrium quantum transport

Juho Lee, Han Seul Kim, and Yong-Hoon Kim*

School of Electrical Engineering, Korea Advanced Institute of Science and Technology (KAIST), Daejeon, 305-701, Korea

*Corresponding author: y.h.kim@kaist.ac.kr

**ABSTRACT:** While the Landauer viewpoint constitutes a modern basis to understand nanoscale electronic transport and to realize first-principles implementations of the non-equilibrium Green's function (NEGF) formalism, seeking an alternative picture could be beneficial for the fundamental understanding and practical calculations of quantum transport processes. Herein, introducing a micro-canonical picture that maps the finite-bias *quantum transport* process to a drain-to-source or multi-electrode *optical excitation*, the multi-space constrained-search density functional theory (MS-DFT) formalism for first-principles electronic structure and quantum transport calculations is developed. Performing MS-DFT calculations for the benzenedithiolate single-molecule junction, it is shown that MS-DFT and standard DFT-NEGF calculations produce practically equivalent electronic and transmission data. Importantly, the variational convergence of "non-equilibrium total energy" within MS-DFT is demonstrated, which should have significant implications for *in operando* studies of nanoscale devices. Establishing a viable alternative to the Landauer viewpoint, the developed formalism should provide valuable atomistic information in the development of next-generation nanodevices.

## 1. INTRODUCTION

Density functional theory (DFT) in the standard form cannot be applied to non-equilibrium quantum electron transport phenomena, thus in the last decade or so the method combining DFT and non-equilibrium Green's function (NEGF) formalism within the Landauer viewpoint has been established as the standard approach for first-principles finite-bias quantum transport calculations.[1, 2] While successful, the DFT-NEGF approach suffers from several shortcomings due to the Landauer framework invoked in its numerical realization. The device is viewed within the Landauer picture as an open system with an external battery as the source of the current flow,[3-5] and to implement this viewpoint one replaces within the self-consistent DFT-NEGF calculations the Hamiltonian and related matrix elements corresponding to the left and right semi-infinite electrode regions with those from two separate infinite bulk DFT calculations. At the fundamental level, the resulting grand canonical formalism makes the total energy, the central object within DFT, ill-defined. Practically, the absence of the variational total energy guideline (and reminding that the electrical current is also not a variational quantity) can be a serious limitation in ensuring the computational convergence of DFT-NEGF self-consistent cycles.[2, 6] Indeed, seeking variational principles for open quantum systems is a long-standing scientific challenge in various contexts.[7, 8]

In this work, we report the development of a multi-space constrained-search formulation of DFT for non-equilibrium electronic structure and quantum transport calculations and its application to a molecular junction. To establish the multi-space constrained-search DFT (MS-DFT) formalism for nanoscale junctions where the grand canonical Landauer picture can be invoked (i.e. a nanoconstriction sandwiched by electron reservoirs), we first adopt the micro-canonical picture and replace the semi-infinite metallic electrodes by finite but sufficiently large metal slabs. Next, we introduce a new conceptual framework that maps the finite-bias *quantum transport* process to the drain-to-source or multi-space electronic *optical excitation*. The resulting MS-DFT provides an alternative framework to the standard DFT-NEGF scheme for first-principles non-equilibrium quantum transport calculations and can be straightforwardly implemented within an existing DFT code. Regarding our initial report on the development of MS-DFT,[9] concern was raised on its validity in that we demonstrated its equivalence with DFT-NEGF only for the vertically-stacked two-dimensional systems involving weak van der Waals interactions. In this and accompanying articles,[10] taking the more established molecular junctions that involve covalent bonding between molecules and electrodes, we show that the quantum transport properties calculated within the standard DFT-NEGF method are faithfully reproduced by the MS-DFT approach. Importantly, the variational convergence of the "non-equilibrium total energy" with respect to the basis-set level within MS-DFT will be demonstrated, which should have significant implications for the first-principles investigations of electrified interfaces for nano-electronic/energy/bio device applications.

## 2. FORMULATION OF MS-DFT

For the nanoscale junctions to which the Landauer picture is applicable (physical or energetic nano-constrictions),[3, 5] we establish the multi-space constrained-search DFT formalism as follows:

*Step 1. Micro-canonical viewpoint.* We first switch from the standard grand-canonical or Landauer picture to the micro-canonical one, in which electrical currents can be viewed as the long-lived discharging of large but finite capacitors. Such an approach was initially explored by Di Ventra and Todorov,[11] but they focused on adopting it within time-dependent DFT to study transient (rather than steady-state) electron dynamics.[12] We will here focus on the steady-state quantum transport problem.

*Step 2. Partitioning.* Next, we divide the junction into the left reservoir ($L$), channel ($C$), and right reservoir ($R$) regions, and trace the spatial origins of a wave function $\Psi$ to $L$ or $C$ or $R$. At the zero-bias limit, together with one global Fermi level, they collectively give the ground-state density $\rho_0(\vec{r}) = \rho_0^L(\vec{r}) + \rho_0^C(\vec{r}) + \rho_0^R(\vec{r})$. The $L/C/R$ partitioning is performed in the spirit of the Landauer picture (**Figure 1**a), and is accordingly similar to that introduced in typical standard DFT-NEGF calculations. Regarding the Landauer framework, we first note that — as emphasized by Landauer himself multiple times — it is a viewpoint and not a specific equation.[5] Especially, it rests upon several physical assumptions that include the presence of a geometrically or energetically narrow nanoconstriction[3, 5] within the channel region $C$ that is sandwiched between the left reservoir $L$ and right reservoir $R$. The $L$ and $R$ reservoirs are individually in local equilibrium with the electrochemical potentials $\mu_L$ and $\mu_R$, respectively, and thus characterized by the Fermi-Dirac functions,

$$f^{L/R}(E) = \frac{1}{1 + \exp[(E - \mu_{L/R})/k_B T]}. \quad (1)$$

It should be cautioned that the $L$ and $R$ "reservoirs" in principle do not correspond to the physical metal "electrodes" in that the interfacial "lead" regions of electrodes that partly accommodate voltage drops (see Figure 1a) should be included within the channel $C$.[3, 5] Within these preconditions that also determine the validity of DFT-NEGF as well as MS-DFT calculations, we now model the electrodes by finite metal slabs and postpone the switching to the grand canonical picture or semi-infinite electrode case until the self-consistent electronic structure calculation is completed (*Step 4*).

*Step 3. Optical analogy.* Finally, we view the quantum *transport* under a finite applied bias voltage $V_b = (\mu_L - \mu_R)/e$ as the *excitation* from the drain reservoir $R$ to the source reservoir $L$, and apply a constrained-search procedure or minimize the total energy to those spatially "excited" states with density $\rho_k$. In establishing the mapping of the transport problem to the optical counterpart, we generalize the variational (time-independent) excited-state DFT, which was formally well-established by Görling[13] and Levy-Nagy,[14] to the $R$-to-$L$ multi-space excitation case. In other words, the role of light in time-independent DFT is played by the external battery in MS-DFT, and it is mathematically embodied by a multi-space constraint.

Then, given the ground state with the total energy $E_0$ and density $\rho_0$, the governing equation of the MS-DFT becomes a constrained search of the total energy minimum of the excited state $k$ with the density $\rho_k(\vec{r}) = \rho_k^L(\vec{r}) + \rho_k^C(\vec{r}) + \rho_k^R(\vec{r})$,

$$E_k = \min_\rho \left\{ \int v(\vec{r})\rho(\vec{r})d^3\vec{r} + F[\rho_k^L, \rho_k^C, \rho_k^R, \rho_0] \right\}$$
$$= \int v(\vec{r})\rho(\vec{r})d^3\vec{r} + F[\rho_k, \rho_0], \quad (2)$$

with the universal functional

$$F[\rho_k, \rho_0] = \min_{\Psi^{L/C/R} \to \rho_k} \langle \Psi^{L/C/R} | \hat{T} + \hat{V}_{ee} | \Psi^{L/C/R} \rangle, \quad (3)$$

where the spatially-resolved $\Psi^{L/C/R}$ are understood to be restricted to the states that satisfy the bias constraint of $eV_b = \mu_R - \mu_L$ and are orthogonal to the first $k$-1 excited states.

For practical calculations, we now invoke the single-electron Kohn-Sham (KS) picture and solve the KS equations,

$$[\hat{h}_{KS}^0 + \Delta v_{H_{XC}}(\vec{r})]\psi_i(\vec{r}) = \varepsilon_i \psi_i(\vec{r}), \quad (4)$$

and obtain $\rho_k$ and $E_k$ by partitioning the $L/C/R$ regions (Figure 1b left panel) and applying the constraint of $eV_b = \mu_L - \mu_R$ (Figure 1b right panel). Here, $\hat{h}^0_{KS}$, $\Delta v_{H_{XC}}(\vec{r}), \psi_i(\vec{r})$, and $\epsilon_i$ represent the ground-state KS Hamiltonian, its bias-induced modification, KS eigenstates, and KS eigenvalues, respectively.

*Step 4. Quantum transport properties.* Recall that within DFT-NEGF the matrix elements of the *L* and *R* reservoir electrode regions are, according to the Landauer viewpoint, replaced by those of separate bulk calculations (scattering boundary conditions) in the process of constructing the self-energy matrix $\mathbf{\Sigma}_{L(R)} = \mathbf{x}_{L(R)}\mathbf{g}_s^{L(R)}\mathbf{x}^\dagger_{L(R)}$, where $\mathbf{x}_{L(R)}$ is the $L-C$ ($C-R$) coupling matrix and $\mathbf{g}_s^{L(R)}$ is the $L$ ($R$) surface Green's function. This replacement in DFT-NEGF directly affects the self-consistent construction of the finite-bias density matrix, or the electron correlation function $\mathbf{G}^n$ (or lesser Green's function $\mathbf{G}^<$) according to

$$\mathbf{G}^n(E) \equiv -i\mathbf{G}^<(E) \approx \mathbf{G}(E)\mathbf{\Gamma}_L(E)\mathbf{G}^\dagger(E)f_L(E) + \mathbf{G}(E)\mathbf{\Gamma}_R(E)\mathbf{G}^\dagger(E)f_R(E), \quad (5)$$

which then, as mentioned earlier, can pose practical difficulties already at the stage of self-consistent electronic structure calculations due to the absence of a variational principle.[2, 6-8]

Within MS-DFT, on the other hand, we complete the self-consistency cycle for the solution of nonequilibrium KS equations without introducing $\mathbf{\Sigma}_{L(R)}$ or the information from separate bulk crystal calculations. After the nonequilibrium electronic structure has become available, we then recover the scattering boundary condition or the Landauer picture as a post-processing step and apply the matrix Green's function formalism[15, 16] to calculate the transmission function

$$T(E; V_b) = Tr[\mathbf{\Gamma}_L(E; V_b)\mathbf{G}(E; V_b)\mathbf{\Gamma}_R(E; V_b)\mathbf{G}^\dagger(E; V_b)], \quad (6)$$

where $\mathbf{G}$ is the retarded Green's function, and $\mathbf{\Gamma}_{L(R)} = i(\mathbf{\Sigma}_{L(R)} - \mathbf{\Sigma}^\dagger_{L(R)})$ is the $L$ ($R$) electrode-induced broadening matrix. The current-bias voltage ($I–V_b$) characteristic is then calculated using the Landauer-Büttiker formula,[1, 2]

$$I(V_b) = \frac{2e}{h}\int_{\mu_L}^{\mu_R} T(E; V_b)[f(E - \mu_R) - f(E - \mu_L)]dE. \quad (7)$$

## 3. IMPLEMENTATION OF MS-DFT

Several comments are in order: First, regarding the transport-to-excitation mapping idea that plays a central role in this work, we emphasize that it was invoked as a viewpoint that replaces the Landauer picture.[3-5] While it was utilized here to formulate the MS-DFT approach, we expect that it could be more generally applicable in treating non-equilibrium quantum transport problems and could be adopted within, e.g., the quantum Monte Carlo approach.

We implemented MS-DFT within the SIESTA code,[17] which is based on the linear combination of atomic orbital formalism and has been extensively employed for the development of DFT-NEGF programs.[18-21] Here, note that the MS-DFT formalism is not restricted to localized atomic-like basis sets and could be generally implemented by constructing Wannier functions.[22] To realize *Step 2*, the *L/C/R* partitioning, we assigned the spatial origins of KS states to *L* or *C* or *R* by comparing their weights within the *L*, *C*, and *R* regions according to

$$\psi_i(\vec{r}) \in \begin{cases} \psi_i^L & \text{if } \int_L |\psi_i(\vec{r})|^2 d^3r > \int_{C/R}|\psi_i(\vec{r})|^2 d^3r, \\ \psi_i^C & \text{if } \int_C |\psi_i(\vec{r})|^2 d^3r > \int_{L/R}|\psi_i(\vec{r})|^2 d^3r, \\ \psi_i^R & \text{if } \int_R |\psi_i(\vec{r})|^2 d^3r > \int_{L/C}|\psi_i(\vec{r})|^2 d^3r. \end{cases} \quad (8)$$

Another important ingredient we need to devise in implementing *Step 3*, the transport-to-excitation mapping, is the rule to determine the occupancy $f_i^C$ associated with the channel region KS states $\psi_i^C$. The assignment of $f^C$ corresponds to the procedure of determining quasi-Fermi levels (QFLs) or the channel electrochemical potential, and $f_i^C$ and $\psi_i^C$ together allow to construct $\rho_k^C(\vec{r})$ according to

$$\rho_k^C(\vec{r}) = \sum_i f_i^C |\psi_i^C(\vec{r})|^2. \quad (9)$$

After testing several possibilities, we determined that the proper occupation rule is to assume that $\psi_i^C$ can be divided into one group originating from *L* and right-traveling with a population function $f_L$ and the other group originating from *R* and left-traveling with a population function $f_R$:

$$f^C(E) \equiv f(E - \mu_C) \rightarrow \begin{cases} f^L(E) & \text{if } \int_L |\psi_i^C(\vec{r})|^2 d\vec{r} > \int_R |\psi_i^C(\vec{r})|^2 d\vec{r}, \\ f^R(E) & \text{if } \int_R |\psi_i^C(\vec{r})|^2 d\vec{r} > \int_L |\psi_i^C(\vec{r})|^2 d\vec{r}, \end{cases}$$

(10)

This rule essentially corresponds to the scheme that was also adopted within DFT-NEGF [at the step of calculating the electron correlation function $\mathbf{G}^n$ according to Equation (5)]. We have recently shown that, given that the *C*-region QFLs can be split into multiple levels in certain situations (**Figure 1a**), maintaining the notion of two separate electrode-originated QFLs is essential in performing MS-DFT calculations.[9]

## 4. RESULTS AND DISCUSSION

We now apply MS-DFT to the molecular junction based on a benzenedithiolate (BDT) molecule (**Figure 2**a), which has been treated by several theoretical groups so can serve as a benchmark system (for details, see Experimental Section and the Supporting Information).[23-28] In Figure 2b, we first show for the BDT junction under $V_b = 0.8$ V the plane-averaged bias-induced electrostatic potential change profiles,

$$\Delta \bar{v}_H(\vec{r}) = \bar{v}_H^V(\vec{r}) - \bar{v}_H^0(\vec{r}), \quad (11)$$

where $\bar{v}_H^V$ and $\bar{v}_H^0$ are the Hartree potentials at the non-equilibrium and equilibrium conditions, respectively, calculated within MS-DFT (red solid lines) and DFT-NEGF (black dashed lines). We indeed find that the linear electrostatic potential drop profile across the molecule obtained from DFT-NEGF is accurately reproduced by MS-DFT, confirming the practical equivalence between the two approaches. We mentioned earlier that the determination of QFLs is a nontrivial matter. For example, instead of the *C*-region $f^C$ occupation rule that maintains the notion of separate *L*- and *R*-originated nonlocal QFLs [equation (10)], enforcing a single averaged electrochemical potential $(\mu_L + \mu_R)/2$ to be $f^C$ resulted in significant errors (blue dotted lines). While the details of the error trend are different, this conclusion is in agreement with our finding from different molecular junction systems.[9]

As one central result of the development of the MS-DFT formalism, we next show in Figure 2c the numerical convergence behavior of the *non-equilibrium* total energy $E_k$ at $V_b = 0.8$ V with the increase of the number of basis-set sizes. It should be emphasized that the total energy becomes an ill-defined quantity once the grand canonical Landauer picture is adopted, so considering the variational convergence of the total energy as in Figure 2c is not possible within DFT-NEGF. However, remaining within the microcanonical viewpoint up to the completion of the finite-bias electronic structure calculation, MS-DFT allows the non-equilibrium total energy to be variationally determined. In contrast to the total energy, electrical current is not a variational quantity.[2, 6] So, as shown together in Figure 2c, the calculated current values exhibit small fluctuations even beyond the level where the total energy has been converged (red squares). Overall, considering the numerical convergences in both the total energy and current, we chose to use the double ζ-plus-polarization-level numerical atomic orbital basis sets in subsequent MS-DFT calculations.

We next examine the quantum transport characteristics of the BDT junction obtained by MS-DFT and DFT-NEGF (for details, see Experimental Section and the Supporting Information). Here, remind again that, in contrast to DFT-NEGF, transport properties are obtained within MS-DFT as a post-processing process after the converged non-equilibrium junction electronic structure is obtained (*Step 4*). In **Figure 3**a, we present the $I$-$V_b$ characteristic of the BDT junction obtained within MS-DFT, which shows the linearly increasing current with the applied voltage. While the calculated conductance is about an order of magnitude larger than the experimental value due to the self-interaction error within LDA,[28, 29] the $I$-$V_b$ data are in good agreement with the DFT-NEGF ones. From the transmission spectra of the BDT junction obtained at $V_b = 0.8$ V, we observe two prominent transmission peaks at –0.85 eV and –1.43 eV below $(\mu_L+\mu_R)/2$, respectively (marked as ① and ② in Figure 3b, respectively). Closely examining the bias-dependent development of the transmission peaks ① and ② that should have originated from the BDT highest occupied molecular orbital (HOMO) and HOMO-1 levels (Figure 3c left panel), one can note that both transmission peaks become narrower and particularly the peak ② intensity decreases with increasing $V_b$. It should be first emphasized that these behaviors are not the artifacts of the MS-DFT method in that, as explicitly shown in Figure 3c left panel for the $V_b = 0.8$ V case, the MS-DFT transmission spectra almost perfectly match the DFT-NEGF counterparts. Indeed, these features can be also found in several previous reports,[27, 28] but no clear explanations were provided.

We now explain the mechanisms of the bias-dependent changes in transmission spectra based on the non-equilibrium KS states, which are uniquely available within MS-DFT. Before demonstrating the advantage of adopting the micro-canonical approach, we first show in Figure 3c right panel the $V_b = 0.8$ V molecule-projected density of state (PDOS) data that can be obtained from DFT-NEGF as well. While the PDOS spectra hint that the left and right S-originated states are split by the applied bias and produce non-transmitting PDOS peaks ③ and ④, the PDOS analysis is apparently limited by providing indirect explanations. To clarify the bias-induced breaking of the degeneracies in the S-originated HOMO and HOMO-1 levels, we show in Figure 3d the three-dimensional contour plots of the MS-DFT-derived KS states corresponding to the conducting PDOS peaks ① and ② as well as the non-conducting PDOS peaks ③ and ④. They then clearly show the delocalized nature with significant weights in the benzene core regions of the KS states ① and ② but the left/right S-localized and disconnected nature of the KS states ③ and ④, providing intuitive explanations of the bias-dependent development of the HOMO and HOMO-1 transmission peaks.

In closing, we comment that there is currently available an analysis method to extract the eigenstates of the molecular projected self-consistent Hamiltonian (MPSH) derived from a DFT-NEGF calculation as a post-processing process.[30] However, the MPSH scheme remains non-rigorous in that one introduces an arbitrarily defined supercell that contains a vacuum and extracts for the cluster model $\Gamma$-point eigenstates. In this regard, we suggest that MS-DFT now provides the formal justification and practical guidelines to take the DFT-NEGF output and properly perform the MPSH post-processing analysis (i.e. take the full *L/C/R* model and the same periodic boundary condition $\vec{k}$-point sampling as in the DFT-NEGF calculation) and even extract the QFLs (i.e. partition the *L/C/R* states and apply the occupation rule as prescribed above). Another promising extension of MS-DFT would be the incorporation of advanced orbital-dependent exchange-correlation energy functionals,[31, 32] which is a non-trivial task once one moves into the NEGF framework.[28] This could resolve the incorrect electrode-channel level alignment problem resulting from self-interaction errors within (semi)local exchange-correlation functionals, finally achieving the quantitative agreement between experimental and theoretical current values.[28, 29]

## 5. CONCLUSION

In summary, we established a novel viewpoint that maps a quantum transport process to a multi-space (from drain to source) "excitation" within the micro-canonical picture. While the viewpoint could be beneficial in a more general context, it particularly allowed us to develop a first-principles quantum transport computational approach based on the MS-DFT framework. The resulting MS-DFT method provides an alternative route to the standard DFT-NEGF scheme for *ab initio* non-equilibrium electronic structure and quantum transport calculations and was straightforwardly implemented within a standard DFT code. To prove the validity of MS-DFT, we then carried out comparative MS-DFT and DFT-NEGF calculations for the BDT single-molecule junction and demonstrated the excellent level of agreements between the two calculations. We thus developed in this work a viable and effective alternative to the DFT-NEGF approach for first-principles quantum transport calculations and additionally pointed out that MS-DFT provides practical guidelines for the post-processing analysis of DFT-NEGF calculation results. Most importantly, the formulation of MS-DFT was achieved by seeking an alternative to the Landauer picture for nanoscale quantum transport, whose significant implication was explicitly displayed *via* the variational convergence of the *non-equilibrium* total energy.

## EXPERIMENTAL SECTION

**DFT calculations.** Based on our earlier works,[16, 29] we modeled the BDT-based molecular junction by locating the S atoms at the FCC hollow sites of the 3 × 3 Au(111) electrode surfaces with the Au-Au gap distance of 9.79 Å. Eight layers of Au atoms were adopted as the electrodes and the outermost three layers were treated as reservoir regions. We checked this setup is sufficient to properly capture the voltage drop across the lead and molecule regions and to support the bias voltage $eV_b$ between the *L* and *R* reservoirs. All calculations were performed within the local density approximation (LDA)[33] by employing double-$\zeta$-plus-polarization-level numerical atomic orbital basis sets and the Troullier-Martins type norm-conserving pseudopotentials.[34] The mesh cutoff of 400 Ry for the real-space integration, and 5 × 5 × 1 Monkhorst-Pack *k*-points grid of the Brillouin zone were sampled. The atomic geometries of the *C* region that includes the lead Au layers were optimized until the total residual forces on each atom were below 0.01 eV/ Å.

**DFT-NEGF and MS-DFT calculations.** We used the TranSIESTA code for DFT-NEGF calculations[19]. For the accurate and quantitative comparison of the DFT-NEGF and MS-DFT calculation results, we applied the identical *L/C/R* partitioning scheme as schematically depicted in Figure 2a and used the same $\mathbf{g}_s^{L(R)}$ matrices. The $\mathbf{g}_s^{L(R)}$ were extracted from separate DFT calculations for three-layer 3 × 3 Au(111) crystals (red and blue shaded area in Figure 2a) with the 5 × 5 $k_\parallel$-point sampling along the surface direction and 10 $k_\perp$-point sampling along the surface-normal direction. The numerical convergence of the calculated transmission spectra with respect to the Au electrode size within DFT-NEGF [35] and MS-DFT is discussed in Supporting Information Figure S1.

## ACKNOWLEDGMENTS

This work was supported by the Nano-Material Technology Development Program (No. 2016M3A7B4024133), Basic Research Program (No. 2017R1A2B3009872), Global Frontier Program (No. 2013M3A6B1078882), and Basic Research Lab Program (No. 2020R1A4A2002806) of the National Research Foundation funded by the Ministry of Science and ICT of Korea. Computational resources were provided by the KISTI Supercomputing Center (KSC-2018-C2-0032).

## AUTHOR CONTRIBUTIONS

Y.-H.K. developed the theoretical framework and oversaw the project. J.L. and H.S.K. implemented the method and J.L. carried out calculations. Y.-H.K. and J.L. analyzed the computational results. Y.-H.K. wrote the manuscript with significant input from J.L.

## REFERENCES

[1]  S. Datta, *Electronic Transport in Mesoscopic Systems*, World Scientific 1995.
[2]  M. Di Ventra, *Electrical Transport in Nanoscale Systems*, Cambridge University Press, Cambridge, UK 2008.
[3]  R. Landauer, Z. Phys. B: Condens. Matter 1987, 68, 217.
[4]  R. Landauer, J. Phys. Condens. Matter 1989, 1, 8099.
[5]  R. Landauer, Physica Scripta 1992, T42, 110.
[6]  Z. Yang, A. Tackett, M. Di Ventra, Phys. Rev. B 2002, 66.
[7]  W. Kohn, Phys. Rev. 1948, 74, 1763.
[8]  B. A. Lippmann, J. Schwinger, Phys. Rev. 1950, 79, 469.
[9]  H. S. Kim, Y.-H. Kim, arXiv preprint 2018, arXiv:1808.03608 [cond.
[10]  J. Lee, H. Yeo, Y. H. Kim, Proc. Natl. Acad. Sci. U. S. A. 2020, 117, 10142.
[11]  M. D. Ventra, T. N. Todorov, J. Phys. Condens. Matter 2004, 16, 8025.
[12]  N. Bushong, N. Sai, M. Di Ventra, Nano Lett. 2005, 5, 2569.
[13]  A. Görling, Phys. Rev. A 1999, 59, 3359.
[14]  M. Levy, Á. Nagy, Phys. Rev. Lett. 1999, 83, 4361.
[15]  Y.-H. Kim, S. S. Jang, Y. H. Jang, W. A. Goddard, 3rd, Phys. Rev. Lett. 2005, 94, 156801.
[16]  Y.-H. Kim, J. Tahir-Kheli, P. A. Schultz, W. A. Goddard, Phys. Rev. B 2006, 73, 235419.
[17]  J. M. Soler, E. Artacho, J. D. Gale, A. García, J. Junquera, P. Ordejón, D. Sánchez-Portal, J. Phys. Condens. Matter 2002, 14, 2745.
[18]  J. Taylor, H. Guo, J. Wang, Phys. Rev. B 2001, 63.
[19]  M. Brandbyge, J.-L. Mozos, P. Ordejón, J. Taylor, K. Stokbro, Phys. Rev. B 2002, 65.
[20]  S.-H. Ke, H. U. Baranger, W. Yang, Phys. Rev. B 2004, 70, 085410.
[21]  A. R. Rocha, S. Sanvito, Phys. Rev. B 2004, 70, 094406.
[22]  N. Marzari, A. A. Mostofi, J. R. Yates, I. Souza, D. Vanderbilt, Rev. Mod. Phys. 2012, 84, 1419.
[23]  K. Stokbro, J. Taylor, M. Brandbyge, J. L. Mozos, P. Ordejón, Comput. Mater. Sci. 2003, 27, 151.
[24]  Y. Xue, M. A. Ratner, Phys. Rev. B 2003, 68, 115407.
[25]  G. C. Liang, A. W. Ghosh, M. Paulsson, S. Datta, Phys. Rev. B 2004, 69, 115302.
[26]  S. H. Ke, H. U. Baranger, W. Yang, J. Chem. Phys. 2005, 122, 074704.
[27]  H. J. Choi, M. L. Cohen, S. G. Louie, Phys. Rev. B 2007, 76, 155420.
[28]  C. Toher, S. Sanvito, Phys. Rev. Lett. 2007, 99, 056801.
[29]  Y.-H. Kim, H. S. Kim, J. Lee, M. Tsutsui, T. Kawai, J. Am. Chem. Soc. 2017, 139, 8286.
[30]  N. Papior, N. Lorente, T. Frederiksen, A. García, M. Brandbyge, Comput. Phys. Commun. 2017, 212, 8.
[31]  M. Städele, J. A. Majewski, P. Vogl, A. Görling, Phys. Rev. Lett. 1997, 79, 2089.
[32]  Y.-H. Kim, M. Städele, R. M. Martin, Phys. Rev. A 1999, 60, 3633.
[33]  D. M. Ceperley, B. J. Alder, Phys. Rev. Lett. 1980, 45, 566.
[34]  N. Troullier, J. L. Martins, Phys. Rev. B 1991, 43, 1993.
[35]  Y.-H. Kim, J. Kor. Phys. Soc. 2008, 52, 1181.

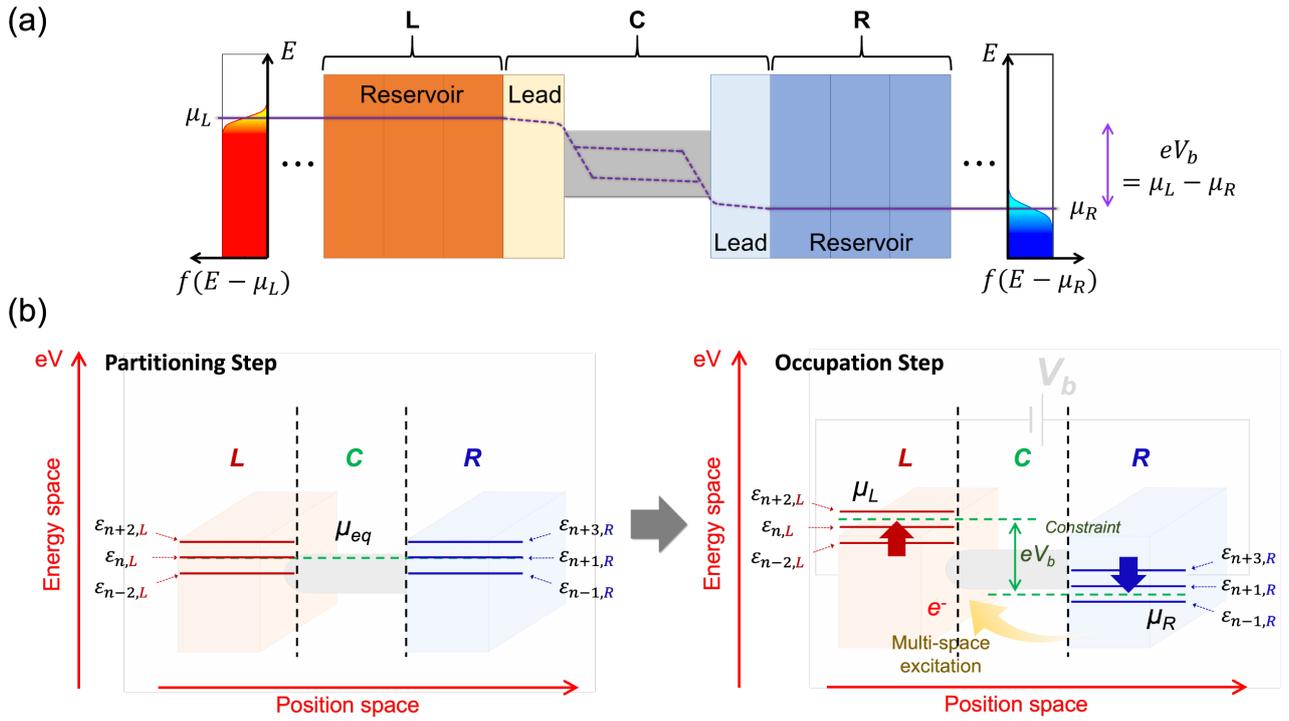

**Figure 1.** (a) Schematic of the Landauer picture for a nanoscale junction under the applied bias voltage $V_b$, where one considers a channel (*C*) sandwiched between the left (*L*) and right (*R*) reservoirs. While the *L* and *R* reservoirs are in local equilibrium with the electrochemical potentials $\mu_L$ and $\mu_R$, respectively, the *C* scattering region is driven into a non-equilibrium state and is characterized by quasi-Fermi (imref) levels that can be split into multiple levels. (b) Schematic of the MS-DFT framework and the computational procedure. The *L*-to-*R* quantum transport is mapped to *R*-to-*L* electron excitation, and it is numerically embodied as a constraint of $eV_b = \mu_L - \mu_R$.

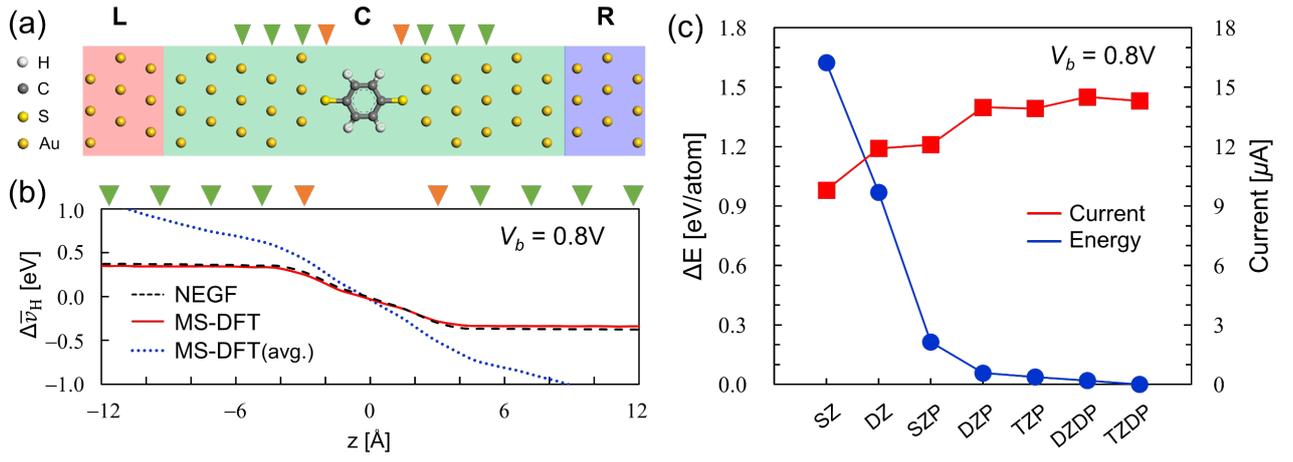

**Figure 2.** (a) The optimized geometry of the BDT junction model. Red, green, and blue boxes indicate the *L*, *C*, and *R* regions, respectively, within the MS-DFT or DFT-NEGF calculations. (b) The plane-averaged electrostatic potential difference $\Delta \bar{v}_H$ calculated for $V_b = 0.8$V within DFT-NEGF (dashed black line), MS-DFT with the occupation rule of Eq. (10) (solid red line), and MS-DFT with a single average occupation rule (dotted blue line). In (a) and (b), the orange and green down triangles respectively indicate the location of the S atoms and Au layers along the transport direction. (c) The numerical convergence behavior of the non-equilibrium total energy (blue filled circles) and current (red filled squares) at $V_b = 0.8$ V with respect to the basis-set level (SZ: single-$\zeta$; DZ: double-$\zeta$; SZP: single-$\zeta$ plus polarization; DZP: double-$\zeta$ plus polarization; TZP: triple-$\zeta$ plus polarization; DZDP: double-$\zeta$ plus double-polarization; TZDP: triple-$\zeta$ plus double-polarization).

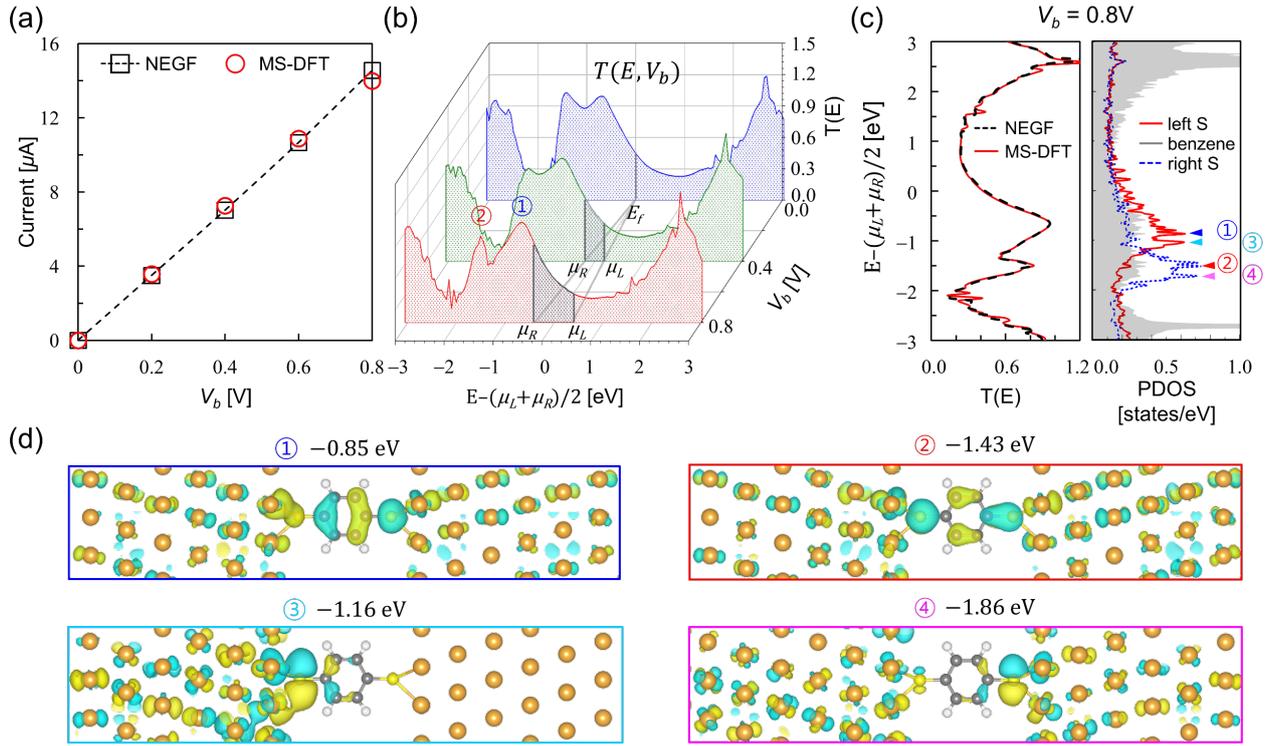

**Figure 3.** (a) Current-bias voltage ($I$-$V_b$) curves of the BDT junction calculated within DFT-NEGF (black open squares, dashed line) and MS-DFT (red open circles). (b) The MS-DFT-calculated transmission spectra for $V_b = 0.0$V (blue), 0.4V (green), and 0.8V (red). The gray shaded areas indicate the transmission spectra within the bias windows $[\mu_L, \mu_R]$ at each $V_b$. The most prominent transmission peaks near $(\mu_L+\mu_R)/2$ are marked as ① and ②. (c) For the $V_b = 0.8$V case, the transmission spectra obtained from DFT-NEGF and MS-DFT are compared (left panel). The molecule PDOS obtained from MS-DFT are shown together (right panel). (d) The three-dimensional contour plots of the MS-DFT-derived KS states that correspond to the PDOS peaks labeled as ①, ②, ③ and ④ in the (c) right panel. In (d), the isosurface level is 0.04 Å$^{-3}$.